\documentclass[12pt]{article}
\usepackage{graphicx}
\usepackage{times}
\usepackage{geometry}
\usepackage[utf8]{inputenc}
\usepackage{enumitem,amssymb}
\usepackage{ragged2e}
\usepackage{wrapfig}
\newlist{thematic}{itemize}{8}
\setlist[thematic]{label=$\square$}
\usepackage{pifont}

\newcommand{\Msun}{$M_{\odot}$}

\newcommand{\mAA}{\AA \,}

\begin{document}
\begin{flushleft}
\huge
Astro2020 Science White Paper \linebreak

Metal Abundances and Depletions in the Neutral Interstellar Medium of Galaxies: the Local Volume as a Laboratory \linebreak
\normalsize

\noindent \textbf{Thematic Areas:} \hspace*{60pt} \linebreak $\square$ Star and Planet Formation \hspace*{20pt}\linebreak
  $\square$  Stars and Stellar Evolution \hspace*{1pt}  \hspace*{40pt} \linebreak
  $\square$    Galaxy Evolution   \hspace*{45pt} \linebreak
  
\textbf{Principal Author:}

Julia Roman-Duval
 \linebreak						
Space Telescope Science Institute
 \linebreak
duval@stsci.edu
 \linebreak
410-338-4351 
 \linebreak
 
\textbf{Co-authors:} \linebreak
Alessandra Aloisi (STScI, aloisi@stsci.edu)\\
Karl Gordon (STScI, kgordon@stsci.edu)\\
Lea Hagen (STScI, lhagen@stsci.edu)\\
Alaina Henry (STScI, ahenry@stsci.edu)\\
Bethan James (STScI, bjames@stsci.edu) \\
Edward B. Jenkins (Princeton University, ebj@astro.princeton.edu)\\
Vianney Lebouteiller (CEA-Saclay, vianney.lebouteiller@cea.fr)\\
Marc Rafelski (STScI, mrafelski@stsci.edu)\\
Kirill Tchernyshyov (The Johns Hopkins University, kirill@jhu.edu)\\
Daniel Welty (STScI, dwelty@stsci.edu)\\
\end{flushleft}

\textbf{Abstract}  (optional):
The comparison of chemical abundances in the neutral gas of galaxies to photospheric abundances of old and young stars, ionized gas abundances, and abundances in galactic halos can trace the chemical enrichment of the universe through cosmic times. In particular, our understanding of chemical enrichment through spectroscopic observations of damped Lyman alpha systems (DLAs) relies on corrections for depletion of metals from the gas to the dust phase. These corrections must be determined in the nearby universe, where both gas-phase abundances and photospheric abundances of young stars recently formed out of the interstellar medium can be measured. Multi-object high-resolution (R$>$50,000) ultraviolet (970-2400 \AA) and optical (300-600 nm) spectroscopy toward massive stars in local volume galaxies (D $<$ 15 Mpc) covering a wide range of metallicities (a few \% solar to solar) and morphological types will provide the abundance and depletion measurements needed to obtain a detailed and comprehensive characterization of the lifecycle of metals in neutral gas and dust in galaxies, thereby observationally addressing important questions about chemical enrichment and galaxy evolution.

\pagebreak

\section{Background and Motivation}

\indent Over a galaxy's lifetime, metals are produced in stars and deposited into the surrounding interstellar medium (ISM). These newly minted metals cycle between different phases of the ISM: some remain in the gaseous phase at different temperatures and pressures, others are locked into dust, and still others are ejected through galactic winds into the circumgalactic medium (CGM), where they can rain back down into the ISM. Meanwhile, a galaxy's fuel for star formation accretes in the form of metal-poor inflows that can dilute its metal content. Dusty environments seed the formation of the next generation of stars, which is of higher metallicity than the preceding one and will eventually initiate the next round of enrichment. This incessant cycle of material between stars, interstellar gas and dust, and galaxy halos fundamentally drives the evolution of galaxies by relentlessly increasing their metal content, altering stellar evolution and the structure of the ISM, and regulating star formation. Although the broad outlines of this cycle are well established, details of its fundamental processes are poorly understood.  From an observational perspective, our understanding is limited because interstellar processes can only be unraveled if they are observed in detail in individual clouds across the various phases of the ISM and in individual stars. \\
\indent Our understanding of how heavy elements are produced and moved by galactic flows relies on the comparison between measurements of chemical abundances in old and young stars, ionized, neutral, and molecular interstellar gas and dust, and halos. For example, the fraction of metals in dust (i.e., the dust-to-metal ratio D/M, derived from the interstellar depletions of individual elements) can be inferred from the comparison of heavy element abundances in young stars with abundances in the interstellar gas they were born from (Jenkins 2009, Jenkins \& Wallerstein 2017, Roman-Duval et al. 2019). Such comparisons can only be performed in the nearby Universe, where stars and interstellar clouds of gas and dust can be resolved. Therefore, our understanding of the metal enrichment of the Universe over cosmic times relies on our ability to constrain the mechanisms responsible for the production, mixing, and transport of metals (e.g., from the ISM to the CGM, or from the gas phase to the dust phase and vice-versa) in nearby galaxies. \\
\indent A concrete example of how critical chemical abundance studies in nearby galaxies are to our comprehension of the chemical enrichment of the universe over cosmic times is the interpretation of abundances in damped Lyman-? systems (DLAs). DLAs are observed over a wide range of redshifts using QSO absorption spectroscopy (e.g. Rafelski et al. 2012). Of particular interest is the ratio of $\alpha$-capture elements (e.g., Si, S, C, O, Mg), produced on short timescales ($\sim$ 10 Myr) by type II SNe, to Fe-peak elements (e.g., Fe, Ni, Cr), produced on longer (1 Gyr) timescales by type Ia SNe. Indeed, a high $\alpha$/Fe ratio in the most metal-poor DLAs, which retain the nucleosynthetic signatures of the first episodes of metal enrichment, may provide evidence for enrichment by Pop III stars (Cooke, R.J. et al. 2017). Abundance measurements in DLAs have to be corrected for the depletion of metals from the gas to the dust phase, particularly at metallicities $>$ 1\% solar. Recent measurements of the fraction of metals in the gas (i.e., interstellar depletions) in the Milky Way (M $=$ 5$\times$10$^{10}$ \Msun), LMC (M $=$ 2.74$\times$10$^9$ \Msun, Z = 0.5 Z$_{\odot}$), and SMC (M $=$ 3.1$\times$10$^8$ \Msun, Z = 0.2 Z$_{\odot}$), however, reveal that 1) volatile elements commonly used as tracers of metallicity (e.g., S, Zn) deplete significantly from the gas to the dust phase, even at the metallicity of the SMC, and 2) there exist large differences in the fraction of metals locked in dust between the Milky Way, LMC, and SMC. These local systems are however the only galaxies where such measurements exist. Without depletion measurements at metallicities lower than that of the SMC, we lack a complete picture of how metals deplete from the gas as a function of metallicity, density, and other environmental factors. Given that the mean metallicity of galaxies at the epoch of the peak cosmic star formation is 5-10\% solar (Madau \& Dickinson 2014), the lack of locally derived interstellar depletions and abundance ratios at low metallicity hinders our understanding of the chemical enrichment of the Universe. Abundances and depletions for key elements (Si, Mg, S, C, O, Fe, Zn) must therefore be derived in nearby galaxies, where the metal content of the gas can be compared to that of young stars recently formed out of the ISM.  \\
\indent Space-based multi-object ($\sim$ 100 fold) ultraviolet (UV, 970-2350 \AA) high-resolution spectroscopy (R $>$ 50,000) and ground-based optical (300-600 nm) medium-resolution (R$>$20,000) multi-object spectroscopy toward individual stars in nearby galaxies (D $<$ 15 Mpc), covering a wide range of metallicities (a few \% solar to solar) and morphological types, will provide the abundance and depletion measurements needed to obtain a detailed and comprehensive characterization of the lifecycle of metals in neutral gas and dust in galaxies, thereby observationally addressing important questions about chemical enrichment and galaxy evolution. While such measurements are possible with current instrumentation in the local group (D $\sim$ 1 Mpc), no such observations exist or are currently possible beyond the local group.


\section{A detailed accounting of metals in gas and dust from UV-optical spectroscopy}

\indent In order to understand how metals deplete from the gas into the dust phase, a detailed census of metals in neutral gas and dust is required. While the metals locked in dust cannot be observed directly, gas-phase abundances can be measured accurately for the key components of dust (Fe, Si, Mg, Ni, C, O) and other metals (Zn, S, Cu) using UV medium to high resolution spectroscopy (R $>$ 50,000, in order to resolve the ISM components and minimize the effects of unresolved saturation). The spectral range required to cover key transitions (e.g., OI $\lambda$976, 1039; Mg II $\lambda$1240, Si II $\lambda$1808, Fe II $\lambda$2249, 2260; C II $\lambda$2325) corresponds to 970-2350 \AA.  The column densities of gas phase metals can be derived from analyzing the UV absorption profiles of various electronic transitions. The atomic hydrogen column densities can be determined from Lorentzian or Voigt (depending on the column density range) profile fitting of the Lyman-$\alpha$ line at 1216 \mAA (Welty et al. 2012, Roman-Duval et al, 2019). Finally, the molecular hydrogen (H$_2$) column densities can be derived from fitting the Lyman H$_2$ transitions (Tumlinson et al. 2002) between 1040 \mAA and 1120 \AA.\\
\indent The photospheric abundances of young stars recently formed out the ISM (O, B, and A stars) can then be used as a proxy for the total (gas + dust) neutral ISM abundances. Photospheric abundances of young stars are typically determined from optical spectroscopy between 3000 and 6000 \mAA (Evans et al., 2007, Bresolin et al., 2006, 2007, Hosek et al. 2014), and Fe abundances require high resolution (R $>$ 20,000, Venn et al, 2003, Kaufer et al., 2004).  Once the total (gas + dust) and gas-phase abundances in the neutral ISM are known, the fraction of metals in the gas (i.e., interstellar depletions) and in the dust can be derived.

\section{Interstellar depletions in the Milky Way and Magellanic Clouds and applications to the distant Universe}

\indent The Milky Way (MW), LMC, and SMC are the only galaxies where interstellar depletions, and thus D/M, have been measured in neutral gas with the method outlined above (Jenkins 2009, Tchernyshyov et al. 2015, Jenkins \& Wallerstein 2017, Roman-Duval et al., 2019). In these galaxies, the fraction of metals in the gas-phase decreases with increasing density (Figure 1a), indicating that gas-phase metals accrete onto dust grains more rapidly as the gas column density increases. The depletion patterns observed locally have important consequences for interpreting metallicity measurements in DLAs. Volatile elements (e.g., S, Zn) are typically assumed to suffer no depletion effects and are therefore commonly used as tracers of metallicity in DLAs (e.g. De Cia et al. 2016). For instance, a constant [Si/S] in DLAs is commonly interpreted as a lack of depletion in Si from the gas, given the assumption that Si should deplete from the gas faster than S, if it does deplete at all (Rafelski et al. 2012, Berg et al., 2015). However, in the Milky Way, LMC, and SMC, S does deplete from the gas as the column density increases, by up to 0.8 dex (Figure 1a). Yet, while [Si/S] shows some decrease with increasing [S/H] in the MW, the [Si/S] ratio in the LMC and SMC is relatively constant, because Si and S deplete at approximately the same rate (Figure 1b). Assuming no depletion in Si based on a constant [Si/S] ratio (De Cia et al. 2018) could therefore lead to a significant underestimation of the $\alpha$-element metallicity in DLAs. Accounting for depletion of volatile elements in DLAs requires a locally derived calibration of this effect as a function of total metallicity and hydrogen column density.\\

\begin{figure*}[h]
\includegraphics[width=9.7cm]{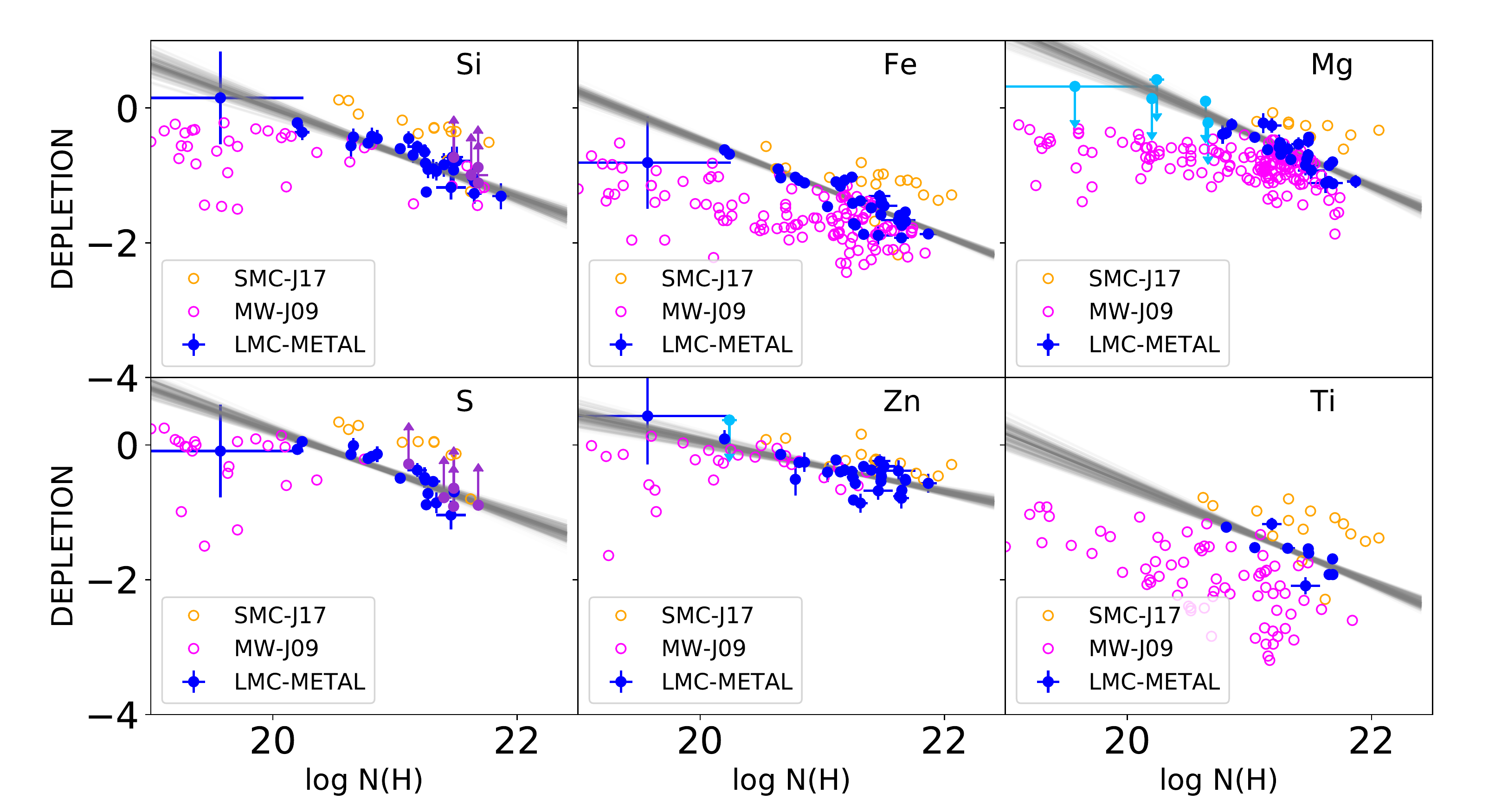}
\includegraphics[width=6.3cm]{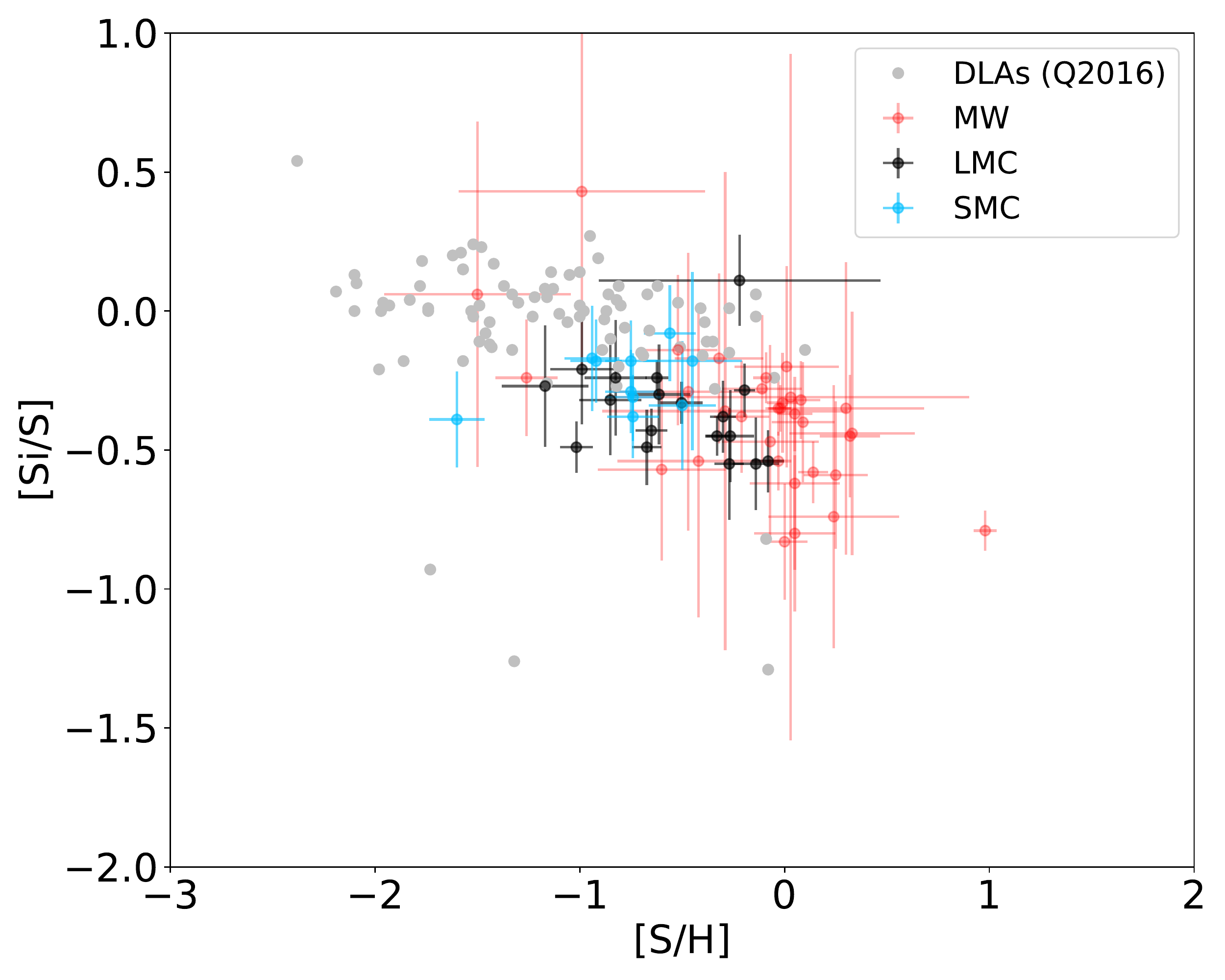}
\caption{(Left) Depletions (log gas-phase fractions) of different elements as a function of hydrogen column density, in the MW (magenta), LMC (blue), and SMC (orange). (Right) Si/S ratio in neutral gas as a function of gas-phase abundance of S in the MW (red), LMC (black), SMC (blue), and DLAs (gray, Quiret et al., 2016). The data are from Jenkins (2009, MW), Roman-Duval et al. (2019, LMC), and Jenkins \& Wallerstein (2017, SMC) and references therein.  }
\end{figure*}

\indent Furthermore, the abundance of Fe-peak elements is best measured directly with Fe, because Zn behaves like an $\alpha$-process element (Ernandes et al. 2018), with [Zn/Fe] being enhanced in some stellar populations in a metallicity-dependent way (da Silveira et al. 2018, Duffau, S. et al. 2017), likely due to complexities in the production mechanism of Zn (Woosley, S. E. \& Weaver, T. A. 1995, Umeda \& Nomoto 2002). However, Fe is significantly depleted, even in the lowest metallicity DLAs, and measuring the $\alpha$/Fe ratio in DLAs therefore requires an estimation of the Fe depletion, which would have to rely on locally derived calibrations of depletions as a function of abundances of other elements and hydrogen column density.

\begin{wrapfigure}{L}{0.5\textwidth}
\includegraphics[width=8cm]{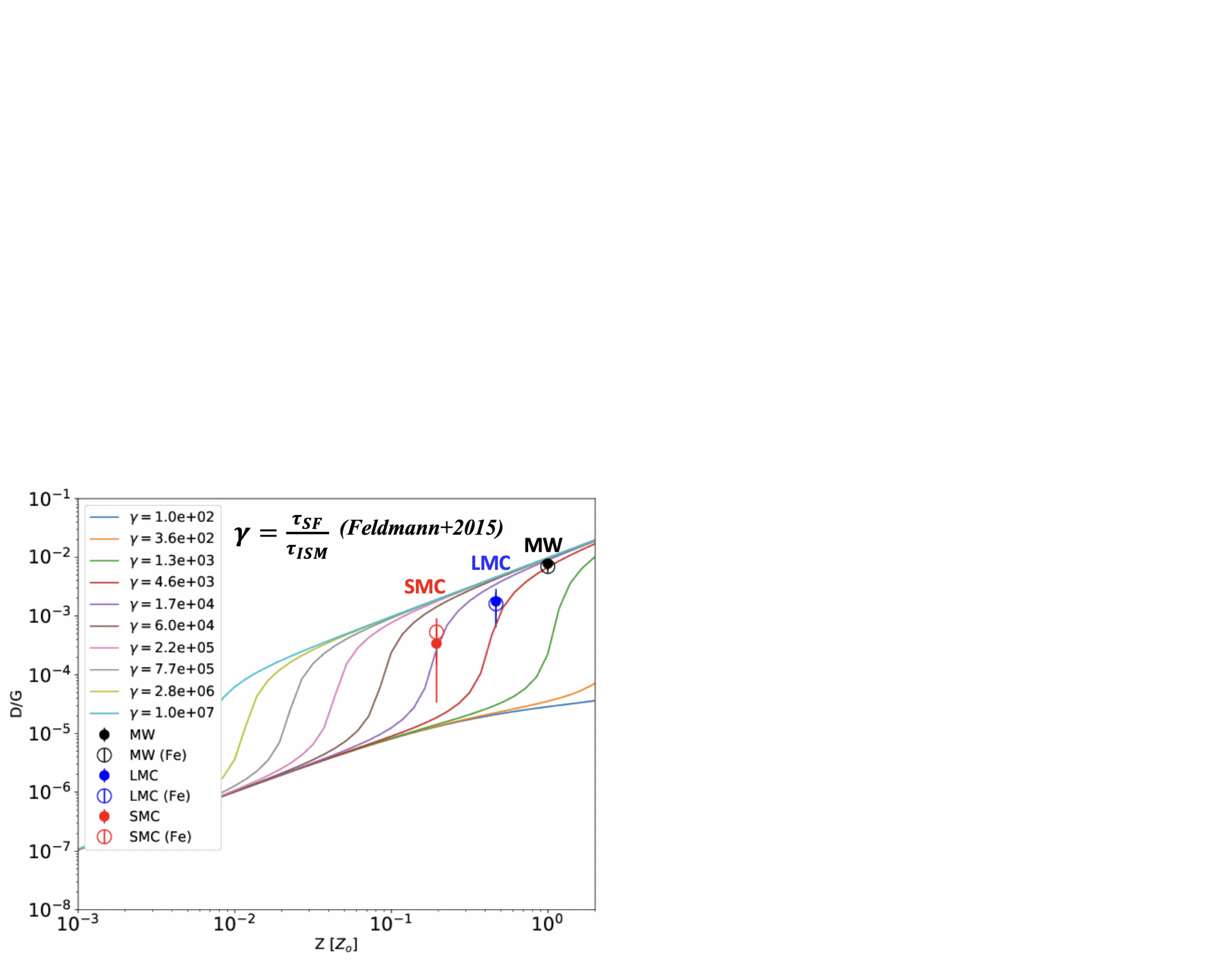}
\caption{Dust-to-gas ratio (D/G) derived from depletion measurements at N(H) = 10$^{21}$ cm$^{-2}$, as a function of metallicity in the MW (black), LMC (blue), and SMC (red). The different model tracks from Feldmann et al. (2015) correspond to different ratios $\gamma$ of the star-formation timescale to the dust formation timescale in the ISM, $\tau_{\mathrm{ISM}}$.}
\end{wrapfigure}


\indent While the fraction of metals in gas and dust depends on the hydrogen column density, it also varies substantially with (total) metallicity: the fraction of metals in dust is 0.2 dex lower in the LMC than in the MW, and 0.5 dex lower in the SMC than in the MW for a given hydrogen column density (Figure 1a). Indeed, chemical evolution models (e.g., Asano et al. 2013, Feldmann et al. 2015) predict that at high metallicity, the main source of dust is growth in the ISM (high D/M), while at low metallicity, galaxies become dominated by star dust (low D/M), resulting in a non-linear relationship between metallicity and D/M and therefore between metallicity and the dust abundance (dust-to-gas ratio D/G $=$ Z$\times$D/M), as shown in Figure 2. The transition between these two regimes occurs at a critical metallicity close to that of the SMC, set by the competition between dust growth in the ISM and the dilution of dust via dust-poor gas inflows. An important consequence of this effect is that the variations in D/M between the MW, LMC and SMC offset almost exactly the differences in metallicities (total, gas + dust), leading to the surprising finding that the gas-phase metallicities between these three drastically different systems are about equal to each other for equivalent hydrogen column densities (Figure 12 of Roman-Duval et al. 2019), and follow the upper metallicity outline of gas-phase metallicities in DLAs. As a result, it would be impossible to distinguish between an SMC-like system or MW-like system in DLA samples based on gas-phase metallicity alone, without comprehensive metallicity-dependent calibrations of how the depletions of different elements correlate with each other and with hydrogen column density. Meanwhile, depletions have not been measured outside the Milky Way, LMC, and SMC, and the behavior of D/M and D/G at low metallicity is therefore unconstrained observationally. Given the differences between the depletions behaviors of the Milky Way, LMC, and SMC (both observed and predicted by chemical evolution models), it is critical for us to study the depletion behaviors of other galaxies with different metallicities and morphologies in order to obtain a better understanding of how the depletion calibrations may vary in a more diverse collection of systems with different chemical evolutions.\\
\indent Thus, in order to understand the metallicity and $\alpha$-enhancement evolution of the Universe through DLAs,, we must first understand how metals deplete from the gas as a function of metallicity and column density. Since many DLAs have metallicities below 1/100 solar, and since the mean metallicity of galaxies at the peak of cosmic star formation was 5-10\% solar (Madau \& Dickinson 2004), this understanding can only be built by obtaining large samples of interstellar depletion measurements in multiple metal-poor galaxies in the nearby universe (D $<$ 15 Mpc), where stellar and ISM abundances can be compared.

\section{Needed: Depletions from UV-optical MOS to 15 Mpc}

\indent In order to understand how metals deplete from the neutral gas to the dust phase, large samples of interstellar depletions covering a range of metallicities, particularly below 20\% solar and down to a few \% solar, are required. In the local group (D $\sim$ 1 Mpc), there are only a couple of dwarf galaxies with metallicities below 10\% solar (Sextans-A and B, WLM). Therefore, we must obtain abundances and depletions to further distances (15 Mpc), where the number of galaxies with metallicities lower than 10\% solar grows to 15 or so (Lisenfeld \& Ferrara 1998, James et al. 2017), including the most metal-poor galaxies in the nearby universe, Leo-P (McQuinn et al. 2015, Z = 0.03 Z$_{\odot}$, D = 1.7 Mpc), DDO 68 (Izotov \& Thuan 2007, Sacchi et al. 2016, Z = 0.03 Z$_{\odot}$, D = 12.6 Mpc), and AGC 198691 ("Leoncino", Hirschauer et al. 2016, Z = 0.02 Z$_{\odot}$, D = 7.7 Mpc). These blue compact dwarf galaxies have star formation rates and ionizing radiation fields that are more representative of high-redshift galaxies ($>$10 \Msun/yr) than low metallicity dwarf irregular galaxies ($<$1 M \Msun/yr) in the local group, thus giving us the opportunity to understand the physical processes inherent to chemical enrichment on the scales of molecular clouds and stars. \\
\indent Depletion measurements require 1) medium to high resolution UV spectroscopy (R $>$ 50,000) of UV bright stars with S/N $>$ 20 in the range 970-2350 \mAA (for the ISM gas phase abundances of a variety of metals, Fe, Si, Mg, O, C, Ni, S, Zn, as well as the HI and H$_2$ column densities), and 2) medium resolution optical spectroscopy with S/N $>$ 50 toward  young O, B, and A stars for photospheric abundances, a proxy for total (gas + dust) ISM abundances. The catalogs of UV bright massive stars suitable for UV spectroscopy and the OBA stars suitable for optical spectroscopy should be identified in multi-band imaging with HST or other future space telescopes (e.g., LEGUS survey, Calzetti et al. 2015).\\
\indent Currently, the most powerful instruments available, the COS UV spectrograph onboard the Hubble Space Telescope and the UVES spectrograph at the VLT, can barely reach massive stars in low metallicity galaxies in the Local Group (e.g, WLM, IC1613, Sextans-A, NGC 3109). ISM gas phase abundances from COS spectroscopy exist toward less than a handful of massive stars in those galaxies, due to the prohibitive exposure times ($\sim$ 10 orbits per star to reach S/N of 15 for the brightest stars in the FUV in Sextans-A at 1.3 Mpc). Beyond the Local Group, no such observations exist or are currently possible due to lack of sensitivity. Photospheric abundances of $\alpha$ elements (Si, S, C, N, O, Mg) from the VLT FORS2 instrument exist for these galaxies (Bresolin et al., 2006, 2007, Evans et al., 2007, Hosek et al. 2014), but Fe abundances require UVES' high-resolution, are expensive to achieve ($>$ 15h/star at 1.3 Mpc), and are therefore only available for a small subset of local group low metallicity galaxies (Venn et al. 2003, Kaufer et al. 2004). \\
\indent A UV spectrograph on a space-telescope with a 15m mirror and R$>$50,000 capability could achieve S/N$>$ 20 toward an O5V star with E(B-V) $=$ 0.05 at D $=$ 15 Mpc in 250h (FUV flux 5$\times$10$^{-18}$ erg cm$^{-2}$ s$^{-1}$ \AA$^{-1}$). Obviously, this precludes single object observations, thus motivating the need for multi-object spectroscopy (MOS) in the UV, with the ability to observe tens of objects simultaneously (the typical number of UV bright stars suitable for spectroscopy in nearby galaxies). In the optical, a 30m class telescope could reach S/N $=$ 50 with R$\sim$15,000 toward massive stars at 15 Mpc in 100h (Puech et al. 2018), and optical MOS capability is therefore a requirement.

\textbf{References}

Berg, T.A.M., et al., 2015, MNRAS, 452, 4326\\
Bresolin et al., 2006, ApJ, 648, 1007\\
Bresolin et al., 2007, ApJ, 671, 2028\\
Calzetti, D., et al., 2015, AJ, 149, 51\\
Cooke, R.J., et al., 2017, MNRAS, 467, 802\\
da Silveira., C.R., et al., 2018, A\&A, 614, A149\\
De Cia, A., et al., 2016, A\&A, 596, A97\\
De Cia, A., et al., 2018, A\&A, 613, L2\\
Duffau, S., et al., 2017, A\&A, 604, A128\\
Ernandes, H., et al., 2018, A\&A, 616, A18\\
Evans et al., 2007, ApJ, 659, 1198\\
Feldmann, 2015, MNRAS, 449, 3274\\
Hirschauer, A.S., et al., 2016, ApJ, 822, 108\\
Hosek, M.W., et al, 2014, ApJ, 785, 151\\
Izotov, Y. I., \& Thuan, T. X., 2007, ApJ, 665, 1115\\
James, B.., et al., 2017, MNRAS, 465, 3977\\
Jenkins, E.B., 2009, ApJ, 700. 1299\\
Jenkins, E.B., \& Wallertein, G., 2017, ApJ, 838, 85\\
Kaufer et al, 2004, AJ, 127, 2723\\
Lisenfeld, U., \& Ferrara, A., 1998, ApJ, 496, 145\\
McQuinn, K.B.W., et al., 2015, ApJL, 815, L17\\
Puech, M., et al., 2018, SPIE, 10702, 107028R\\
Quiret et al. 2016, MNRAS, 458, 4074\\
Rafelski, M., et al., 2012, ApJ, 755, 89\\
Roman-Duval, J., et al., 2019, ApJ, 871, 151\\
Sacchi, E., et al., 2016, ApJ, 830,3\\
Tchernyshyov, K., et al., 2015, ApJ, 811, 78\\
Tumlinson, J., et al., 2002, ApJ, 566, 857\\
Venn et al., 2003, AJ, 126, 1326\\
Welty, D.E., \& Crowther, 2010, MNRAS, 404, 1321\\
Welty, D.E., et al., 2012, ApJ, 745, 173\\

\end{document}